\pdfoutput=1
\def\compileforpublish{1}
\def\isaccepted{1}

\documentclass[letterpaper, 10 pt, conference]{ieeeconf}  %

\IEEEoverridecommandlockouts                              %

\usepackage{tikz,amsmath, amssymb,bm,color,graphicx,siunitx}
\allowdisplaybreaks
\usepackage{pdflscape}
\usepackage{url}
\usepackage{array,booktabs,colortbl}
\usepackage{siunitx} 
\usepackage{tikz}
\usepackage{longtable}
\usepackage{multicol}
\usepackage{multirow}
\usepackage{threeparttable}
\usepackage{capt-of}
\usepackage{pgfplotstable}  
\usepackage{csquotes}
\usepackage{wasysym}
\let\labelindent\relax
\usepackage{enumitem}
\usepackage[english]{babel}
\usepackage{tubscolors}
\usepackage[pdfusetitle, pdfauthor={Stolte et al., TU Braunschweig}]{hyperref}
\usepackage[style=ieee,
			backend=bibtex,
			minnames=1,
			maxcitenames=2,
			maxbibnames=99,
			doi=false,
			isbn=false,
			url=false,
			natbib=true,
			clearlang=true,
			date=year
			]
			{biblatex}
\DeclareRedundantLanguages{english,german,french}{english,german,ngerman,french}
\AtEveryBibitem{\clearfield{month}}

\usetikzlibrary{arrows, shapes, positioning, shapes.geometric, decorations.markings, calc, fit, fadings, shadows}
\usetikzlibrary{scopes}
\usetikzlibrary{arrows.meta}
\usetikzlibrary{patterns}
\usetikzlibrary{automata}
\pgfplotsset{compat=newest}
\pgfplotsset{plot coordinates/math parser=false}

\definecolor{lightgrey}{gray}{0.9}
\definecolor{wheelColor}{named}{tuBlack}%
\definecolor{contourColor}{named}{tuRed}%
\definecolor{longForceColor}{named}{tuBlueMedium80}%
\definecolor{latForceColor}{named}{tuBlueMedium80}%

\definecolor{refPosColor}{named}{tuRed}%
\colorlet{refPosColor}{tuRed!80}
\definecolor{refYawColor}{named}{tuRed}%
\colorlet{refYawColor}{tuRed!80}
\definecolor{refPathColor}{named}{tuRed}%
\colorlet{refPathColor}{tuRed!60}

\definecolor{actPosColor}{named}{tuBlueMedium80}%
\definecolor{actPathColor}{named}{tuBlueMedium60}%

\definecolor{deviationColor}{named}{tuBlueMedium40}%

\newcommand{\eg}{{e.\,g.}\ }
\newcommand{\ia}{{i.\,a.}\ }
\newcommand{\ie}{{i.\,e.}\ }
\newcommand{\cf}{{c.\,f.}\ }

\newcommand{\figureref}[1]{Figure~\ref{#1}}
\newcommand{\sectionref}[1]{Section~\ref{#1}}

\newcommand{\vc}[1]{\mathbf{#1}}
\newcommand{\mx}[1]{{\bm{#1}}} 
\newcommand{\vcs}[1]{\mathbf{\bm{#1}}} 
\newcommand{\cg}{{CG}\ }

\makeatletter
\newcounter{IEEE@bibentries}
\renewcommand\IEEEtriggeratref[1]{%
	\renewbibmacro{finentry}{%
		\stepcounter{IEEE@bibentries}%
		\ifthenelse{\equal{\value{IEEE@bibentries}}{#1}}
		{\finentry\@IEEEtriggercmd}
		{\finentry}%
	}%
}
\makeatother

\ifx \isaccepted \undefined
\newcommand\copyrighttext{%
	\footnotesize \centering This work has been submitted to the IEEE for possible publication.\\ Copyright may be transferred without notice, after which this version may no longer be accessible.}
\else
\newcommand\copyrighttext{%
	\footnotesize \parbox[t]{.11\textwidth}{\copyright{} \the\year~IEEE.} \parbox[t]{.89\textwidth}{Personal use of this material is permitted. Permission from IEEE must be obtained for all other uses, in any current or future media, including reprinting/republishing this material for advertising or promotional purposes, creating new collective works, for resale or redistribution to servers or lists, or reuse of any copyrighted component of this work in other works.}}
\fi
\newcommand\copyrightnotice{%
	\ifx \compileforpublish \undefined
	\else
	\begin{tikzpicture}[remember picture,overlay]
		\node[anchor=south,yshift=10.5pt] at (current page.south) {\parbox{\dimexpr\textwidth-\fboxsep-\fboxrule\relax}{\copyrighttext}};
	\end{tikzpicture}%
	\fi
}

\definecolor{mn}{RGB}{255,127,0}
\definecolor{ts}{RGB}{0,0,255}
\definecolor{mr}{RGB}{190,0,80}

\title{\LARGE \bf
	Investigating Functional Redundancies in the Context of Vehicle Automation -- A Trajectory Tracking Perspective*
}

\author{Torben Stolte$^{1}$, Tianyu Liao$^{1}$, Matthias Nee$^{1}$, Marcus Nolte$^{1}$, and Markus Maurer$^{1}$%
\thanks{\hspace{-1em}*This work is part of the DFG Research Unit Controlling Concurrent Change, funding number FOR 1800.}
\thanks{\hspace{-1em}$^{1}$The authors are with the Institute of Control Engineering at Technische Universit\"at Braunschweig, 38106 Braunschweig, Germany. 
        {\tt\small \{stolte, nolte, maurer\}@ifr.ing.tu-bs.de}}%
}

\begin{document}

\maketitle
\thispagestyle{empty}
\pagestyle{empty}

\begin{abstract}%
	Level~3+ automated driving implies highest safety demands for the entire vehicle automation functionality.  
	For the part of trajectory tracking, functional redundancies among all available actuators provide an opportunity to reduce safety requirements for single actuators. 
	Yet, the exploitation of functional redundancies must be well argued if employed in a safety concept as physical limits can be reached. 
	In this paper, we want to examine from a trajectory tracking perspective whether such a concept can be used. 
	For this, we present a model predictive fault-tolerant trajectory tracking approach for over-actuated vehicles featuring wheel individual all-wheel drive, brakes, and steering.
	Applying this approach exemplarily demonstrates for a selected reference trajectory
	that degradations such as missing or undesired wheel torques as well as reduced steering dynamics can be compensated. 
	Degradations at the physical actuator limits lead to significant deviations from the reference trajectory while small constant steering angles are partially critical.
\end{abstract}%

\section{Introduction}
\label{sec:introduction}

In automated vehicles according to SAE level\,3+~\cite{sae_2016}, the driver's role as fallback in conventional vehicles is transferred to the technical system (in level 3 until hand-over to the driver). 
Hence, ensuring safety is of utmost importance. 
Compared to conventional vehicles, this results in significantly increased safety requirements for all parts of the vehicle automation system which in turn lead to elaborate technical implementations. 

On the actuator level, specifically, each actuator must fulfill fail-operational degradation regimes~\cite{stolte_2016} if not otherwise argued. 
All actuators contribute to both lateral and longitudinal dynamics as partially exploited in conventional vehicles to enhance vehicle agility and safety, \eg torque vectoring or electronic stability control. 
Hence, the use of these functional redundancies among the actuators appears promising in order to reduce safety requirements on the actuator level. 

From a functional perspective, following \citet{matthaei_2015}, fault-tolerant trajectory tracking is the first reaction to a degradation on operational level. 
The trajectory tracking keeps the deviation from the reference trajectory in terms of a temporal sequence of reference poses (position plus heading) in defined boundaries.
Still on the same level, trajectory generation prevents non-drivable trajectories by re-planning the reference trajectory under consideration of the actuator degradation \cite{nolte_2017}. 
On tactical level, the targeted vehicle behavior must adapt to the degradation, for instance by decreasing speed or even stopping the vehicle on a hard shoulder. 
Last but not least, the planned route can be changed on strategical level, \eg driving to a garage for repair or avoiding steep or curvy roads.

In the following we focus on the part of fault-tolerant trajectory tracking. 
Furthermore, we concentrate on an over-actuated actuator topology featuring wheel-individual brakes, drives, and steering, \ie the topology with the highest force potential~\cite{jonasson_2010a}. 
Still, utilization of functional redundancies in a safety concept for automated vehicles has not been subject to comprehensive investigation. 
For the part of trajectory tracking, the suitability is determined by aspects such as the tolerable degradation types, the actuator topology as well as the vehicle's functional range (allowed velocities, lateral accelerations, etc.) and its operational design domain~\cite{sae_2016}.
The operational design domain defines \ia friction coefficient ranges through the admissible road and weather conditions and the accepted tracking error for each possible operational scenario (\cf~\cite{stolte_2017} for definition).

Several publications present trajectory tracking approaches for the selected actuator topology demonstrating its benefits from a vehicle dynamics perspective. 
Most of these approaches target the vehicle's lateral dynamics, \eg \cite{fredriksson_2004,knobel_2006,ono_2006,orend_2005,orend_2005a}. 
One exception is the work of \citet{park_2015,park_2017} who track a trajectory in terms of a temporal sequence of reference poses. 
Additionally, trajectory tracking approaches, which provide fault-tolerance against actuator degradations, consider selected degradation types of single actuators only  \cite{hac_2006,yang_2008,yang_2010,wang_2011,wang_2012,wang_2013}, target lateral dynamics or path tracking \cite{hac_2006,yang_2008,yang_2010,li_2013}, or do not demonstrate limitations of their approaches~\cite{hac_2006,yang_2008,yang_2010,wang_2011,wang_2012,wang_2013,li_2013,moseberg_2016}.

In order to investigate the potential as well as limitations of functional redundancies, we implemented a model predictive control (MPC) approach. 
The approach is able to track a temporal sequence of references poses and can incorporate the effects of various types of actuator degradations. 
From an architectural perspective, we assume the remaining system parts as non-reactive to actuator degradations for this paper. 
In \sectionref{sec:MPC}, we outline the control scheme which is subsequently evaluated in \sectionref{sec:evaluation} at the example of a dynamic reference trajectory in presence of various degradations.

\section{Fault-Tolerant Trajectory Tracking}
\label{sec:MPC}
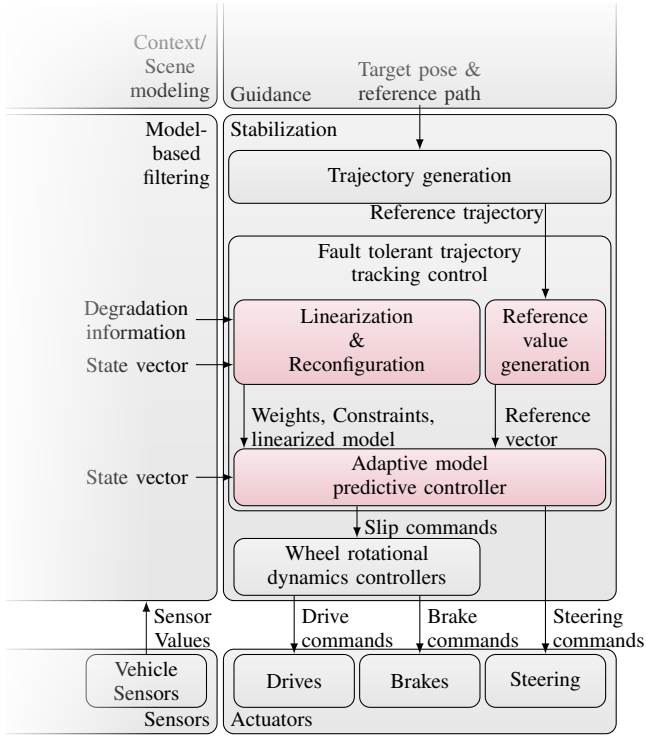
\begin{figure}[t]
	\centering
	 \tikzstyle{shadedGrayLight} = [top color=tuBlack!5, bottom color=tuBlack!10, draw=tuBlack!30]
 
 \tikzstyle{outerblock} = [draw, shadedGrayLight, rounded corners,minimum width=\blockwidth]
 \tikzstyle {archblock} = [outerblock, minimum height=2.5em, align=center]

\begin{tikzpicture}

	\node [] (topright)   at (3.25cm, 14em){};
	\node [] (bottomleft) at (-5.5cm,-14em){};
	\clip[](bottomleft) rectangle  (topright);
		
	\footnotesize
	\pgfdeclarelayer{background}
	\pgfdeclarelayer{foreground}
	\pgfsetlayers{background,main,foreground}
		\tikzset{>=latex}
		
		\tikzstyle{FTMPC}=[draw, top color=tuRed!10, bottom color=tuRed!25,minimum height=2.5em, minimum width=1.5cm,rounded corners,align =center]
 	 	\tikzstyle{others}=[draw=tuBlack, top color=tuBlack!5, bottom color=tuBlack!10, minimum height=2.5em,rounded corners,minimum width=\blockwidth,align=center]
		\tikzstyle{ann} = [above right,align=left,yshift=1em]
		\def\yshiftAMPC{3.}
		\def\edgedist{2}
		\def\yblockdistance{3em}
		\def\xblockdistance{.25em}
		\def\blockwidth{1.6cm}
		\def\xgap{.25em}
		\def\ygap{.25em}
		\node[others,minimum width=3*\blockwidth+2*\xblockdistance+2*\xgap, minimum height = 13em] (FTTT){};
		\node[align=center,below] at (FTTT.north)(drop){Fault tolerant trajectory\\tracking control};
		\node [others,anchor=south west,minimum width= 2*\blockwidth+\xblockdistance] at ($(FTTT.south west)+(\xgap,-4*\ygap-\yblockdistance)$)(LSR){Wheel rotational\\dynamics controllers} ;
	 	\node (AMPC) [FTMPC,minimum width =3*\blockwidth+2*\xblockdistance, anchor=south west] at ($(LSR.north west)+(0,.5*\yblockdistance)$) {Adaptive model\\predictive controller};
			
	 	\node [FTMPC, minimum height=4em,anchor =south east,align=center,minimum width= \blockwidth] 	at ($(AMPC.north east)+(0,1*\yblockdistance)$) (STJ) {Reference\\value\\generation};
	 	\node [FTMPC, minimum height=4em,anchor=south west,align=center,minimum width= 2*\blockwidth+\xblockdistance] 	at ($(AMPC.north west)+(0,\yblockdistance)$)(RKF) {Linearization\\\&\\Reconfiguration};
	
		\node [others ,minimum width=3*\blockwidth+2*\xblockdistance+2*\xgap,,anchor=south west] at ($(FTTT.north west)+(0,.5*\yblockdistance)$) (trajgen) {Trajectory generation};
		
	 	\node [others, anchor=north west] at ($(LSR.south west)+(0,-2/3*\yblockdistance-3*\ygap)$)(drives){Drives} ;
		\node [others, anchor=north east] at ($(LSR.south east)+(0,-2/3*\yblockdistance-3*\ygap)$)(brakes){Brakes} ;
		\node [others, anchor=south east] at (AMPC.south east|-brakes.south east)(steering){Steering} ;
		\node [others,anchor=north east] at ($(LSR.south west)+(-5*\xgap,-2/3*\yblockdistance-3*\ygap)$)(vehiclesensors){Vehicle\\Sensors};

	  	\draw [->] ($(RKF.south west)+(2*\xblockdistance,0)$) -- node [right,align=left,yshift=-2*\ygap] {Weights, Constraints,\\linearized model}  ($(AMPC.north -| RKF.south west)+(2*\xblockdistance,0)$);
	  	\draw [->] ($(STJ.south west)+(2*\xblockdistance,0)$) -- node [xshift=.7cm,align=left,yshift=-2*\ygap] {Reference\\vector}  	($(AMPC.north -| STJ.south west)+(2*\xblockdistance,0)$);
	 	
		\draw [->] (AMPC.south-|steering) --  (steering)node [ann,yshift=\ygap] (drop){Steering\\commands} ; 
	  	\draw [->] (AMPC.south-|LSR)  -- (LSR) 	node [ann] {Slip commands}; 
	   	\draw [->] (LSR.south-|drives) -- (drives) 	node [ann,yshift=\ygap] {Drive\\commands}; 
	  	\draw [->] (LSR.south-|brakes) -- (brakes) node [ann,yshift=\ygap] {Brake\\commands}; 
	   	\draw [->] (trajgen.south-|STJ.north) node [ann,xshift=+.3em,yshift=-.7em,below left] {Reference trajectory} -- (STJ) 	;
	  	\draw [->] ($(RKF.west) + (-.5cm,-.3cm)$) node[align=center, anchor=east,xshift=0.cm](drop){State vector} -- ($(RKF.west) + (.0cm,-.3cm)$);
	  	\draw [->] ($(AMPC.west) + (-.5cm,0)$) node[align=center, anchor=east,xshift=0.cm](drop){State vector} -- ($(AMPC.west)$);
	   	\draw [->] ($(RKF.west) + (-.5cm, .3cm)$) node[align=center, anchor=east,xshift=0.cm](drop){Degradation\\information} -- ($(RKF.west) + (.0cm,.3cm)$);
  		\draw [->] ($(trajgen.north) + (0,.6cm)$) node[align=center, anchor=east,yshift=-0.1cm,above](drop){Target pose \&\\reference path} -- ($(trajgen.north)$);
  
 	\begin{pgfonlayer}{background}
		\clip[](bottomleft) rectangle  (topright);
		\node [others,anchor=south west,minimum width= 3*\blockwidth+2*\xblockdistance+4*\xgap,minimum height=23em] at ($(FTTT.south west)-(\xgap,5*\ygap+\yblockdistance)$)(stabilization){} ;
		\node [align=center,anchor= north west] at (stabilization.north west){Stabilization};
		\node [others,anchor=south east,minimum width= 3*\blockwidth+2*\xblockdistance+4*\xgap,minimum height=23em] at ($(stabilization.south west)-(\xgap,0)$)(modelbasedfilter){} ;
 		\node [align=center,anchor=north east] at (modelbasedfilter.north east){Model-\\based\\filtering};
 		\node [others,anchor=south west,minimum width= 3*\blockwidth+2*\xblockdistance+4*\xgap,minimum height=23em] at ($(stabilization.north west)+(0,\ygap)$)(guidance){} ;
 		\node [align=center,anchor=south west] at (guidance.south west){Guidance};
 		\node [others,minimum width= 3*\blockwidth+2*\xblockdistance+4*\xgap,minimum height=23em] at (modelbasedfilter|-guidance)(scenemodeling){} ;
 		\node [align=center,anchor=south east] at (scenemodeling.south east){Context/\\Scene\\modeling};		
 		\node [others,anchor=north,minimum width= 3*\blockwidth+2*\xblockdistance+4*\xgap,minimum height=4em] at ($(stabilization.south)-(0,\yblockdistance-3*\ygap)$)(actuators){} ;
 		\node [align=center, anchor = south west] at (actuators.south west){Actuators};
 		\node [others,anchor=north,minimum width= 3*\blockwidth+2*\xblockdistance+4*\xgap,minimum height=4em] at ($(modelbasedfilter.south)-(0,\yblockdistance-3*\ygap)$)(sensors){} ;
 		\node [align=center, anchor = south east] at (sensors.south east){Sensors};
  		\draw [->] ($(vehiclesensors.north)$) -- (vehiclesensors.north|-modelbasedfilter.south)node[align=center,below right](drop){Sensor\\Values};
	\end{pgfonlayer}
 
	\node [] (topright2)   at ($(topright)+(-7cm,0)$){};
	\node [] (bottomleft2) at (bottomleft){};
   	\node [fill=white, scope fading =east, fit=(topright2)(bottomleft2)](white) {};
   	
 	\node [] (topright2)   at (topright){};
 	\node [] (bottomleft2) at ($(bottomleft)+(0,30em)$){};
    \node [fill=white, scope fading =south, fit=(topright2)(bottomleft2)](white) {};
		
\end{tikzpicture}
	\vspace{-2em}
	\caption{Control scheme presented in this paper (red) embedded in a functional system architecture according to \citet{matthaei_2015}}
	\label{fig:controlstructure}
	\vspace{-1em}
\end{figure}

\copyrightnotice
In \figureref{fig:controlstructure}, the adaptive model predictive control scheme for fault-tolerant trajectory tracking is embedded in a functional system architecture. 
As input, a reference vector containing the target vehicle motion is derived from the reference trajectory in the reference value generation block. 
The adaptive model predictive controller generates steering and slip commands in terms of target steering angles and longitudinal slip, respectively, in order to influence the vehicle dynamics. 
Subsequently, a wheel rotational dynamics controller calculates drive and brake commands as target torques from the target longitudinal slip. 
The recent vehicle state in terms of state vector and degradation information is determined by means of sensor values which are processed within the model-based filtering block. 
This process is not within the focus of this paper. 
Different approaches for degradation detection and isolation are described \ia in \cite{isermann_2005}. 
The state vector serves as input for the model predictive controller as well as as input for the linearization and reconfiguration block which creates a linearized prediction model of the current operating point. 
Concurrently, the degradation information is used to reconfigure the prediction model as well as the weights and constraints of the optimization.

Assuming a discrete non-linear system without feedthrough
\begin{align*}
	\vc{x}_{k+1}&=\vc{f}\left(\vc{x}_k,\vc{u}_k\right),\\
	\vc{y}_{k}  &=\vc{h}\left(\vc{x}_k\right),
\end{align*}
where $\vc{x}$ denotes the state vector, $\vc{y}$ the output vector, $\vc{u}$ the control vector, $k$ a discrete point of time, as well as~$\vc{f}$ and~$\vc{h}$ the transition and output function, 
the fundamental linear optimization problem can be described as 

\vspace{.5em}	
\begin{tabular*}{\columnwidth}{lll}
	\multicolumn{3}{l}{$\underset{\vcs{x},\vcs{u}}{\text{min}}$\quad$J\left(\vcs{y},\vcs{u}\right)$\,,}\\
	{s.\,t.} 	& $\Delta\vc{x}_0$		&$=\vc{0}$\,,\\
				& $\Delta\vc{u}_k$		&$=\vc{u}_{k}-\vc{u}_{0}$\,,\\
				& $\Delta\vc{x}_{k+1}$	&$=\mx{A}^{\mathrm{lin}}_{0}\Delta\vc{x}_{k}+\mx{B}^{\mathrm{lin}}_{0}\Delta\vc{u}_{k}+\vc{r}_0$\,, \\
				& $\vc{y}_{k}$			&$=\mx{C}^{\mathrm{lin}}_{0}\Delta\vc{x}_{k}+\vc{h}_0$\,,\\
				& $\vc{y}_{k}$			&$\in \mathbb{Y}$\,,\\
				& $\vc{u}_{k}$			&$\in \mathbb{U}$ 
\end{tabular*}\vspace{.5em}
with input and output constraints $\mathbb{U}$ and $\mathbb{Y}$, which are convex sets, the linearized state transition matrix $\mx{A}^{\mathrm{lin}}_{0}$, input matrix $\mx{B}^{\mathrm{lin}}_{0}$, and output matrix $\mx{C}^{\mathrm{lin}}_{0}$  
\begin{align*}
	\mx{A}^{\mathrm{lin}}_{0}=\left.\frac{\partial \vc{f}}{\partial \vc{x}}\right|_{\vc{x}_{0},\vc{u}_{0}}\text{, }
	\mx{B}^{\mathrm{lin}}_{0}=\left.\frac{\partial \vc{f}}{\partial \vc{u}}\right|_{\vc{x}_{0},\vc{u}_{0}}\text{, }
	\mx{C}^{\mathrm{lin}}_{0}=\left.\frac{\partial \vc{h}}{\partial \vc{x}}\right|_{\vc{x}_{0},\vc{u}_{0}}\text{, }
\end{align*}\vspace{-1em}
\begin{align*}
	&\Delta\vc{x}_{k}=\vc{x}_{k}-\vc{x}_{0}, 							& &\Delta\vc{u}_{k}=\vc{u}_{k}-\vc{u}_{0},\\
	&\vc{r}_{0}=\vc{f}\left(\vc{x}_{0},\vc{u}_{0}\right)-\vc{x}_{0},	& &\vc{h}_{0}=\vc{h}\left(\vc{x}_{0},\vc{u}_{0}\right)=\vc{y}_0,
\end{align*}
as well as the cost function $J\left(\vc{y},\vc{u}\right)$ 
\begin{align*}
\setlength{\thinmuskip}{0mu}
	\begin{split}
		J\left(\vcs{y},\vcs{u}\right)=
			\sum_{k=0}^{N_P-1}\left[\left(\vc{y}_{k+1}-\vc{y}_{\mathrm{ref},k+1}\right)^T\mx{Q}\left(\vc{y}_{k+1}-\vc{y}_{\mathrm{ref},k+1}\right)\right.\\
					\left.+\left(\vc{u}_{k}-\vc{u}_{\mathrm{ref},k}\right)^T\mx{R}\left(\vc{u}_{k}-\vc{u}_{\mathrm{ref},k}\right)\right],
	\end{split}
\end{align*}
with its input and output weight matrices $\mx{Q}=\text{diag}({\vc{w}_y})$ and $\mx{R}=\text{diag}({\vc{w}_u})$ and the prediction horizon $N_P$. 
An index $(\cdot)_{k=0}$ indicates the actual value, an index $(\cdot)_{\mathrm{ref}}$ a reference value. 
${\vc{w}_y}$ and ${\vc{w}_u}$ are the weight vectors for the elements of the input and output vectors. 

In the following, the fundamental considerations regarding the prediction model are outlined in Subsection~\ref{subsec:mpcscheme}. 
The resulting prediction model is presented in Subsection~\ref{subsec:vehiclemodel} together with the actuator degradation mapping in Subsection~\ref{subsec:failuremapping}.
Table \ref{tab:quantities} summarizes the nomenclature used in the model of the vehicle's motion along a trajectory.

\begin{table}
	\centering
	\caption{Nomenclature}
	\label{tab:quantities}
	\begin{tabular}{m{.29\linewidth}m{.57\linewidth}}
		\toprule
		\multicolumn{2}{l}{Superscripts and Subscripts}\\
		\midrule
		$(\cdot)^a$, $a\in\{\mathrm{O,V,W}\}$	& Global, vehicle, or wheel coordinate system\\
		$(\cdot)_b$, $b\in\{x,y\}$				& Translational or rotational quantity along or around $x$- or $y$-axis of coordinate system\\
		$(\cdot)_i$, $i\in\{\mathrm{f,r}\}$		& Front or rear axle\\
		$(\cdot)_j$, $j\in\{\mathrm{r,l}\}$		& Right or left side\\
		\midrule
		Quantities\\
		\midrule
		$m$										&	Vehicle mass\\
		$J_z$									&	Moment of inertia around $z$-axis in vehicle's \cg\\
		$F^a_{b,ij}$							&	$x$- and $y$-components of the forces at the wheels in wheel or	vehicle coordinates\\
		$F^a_{b}$								&	Resulting forces at the vehicle's \cg\\
		$\vec{F}^{\mathrm{W}}_{ij}$				&	Resulting forces at wheels\\
		$M_z$									&	Resulting yaw moment at the vehicle's \cg\\
		$\vec{v}$								&	Vehicle speed in \cg\\
		$v^{\mathrm{V}}_b$, $\dot v^{\mathrm{V}}_{b}$					&	Longitudinal and lateral velocity and acceleration of vehicle\\
		$v^a_{b,ij}$							&	Longitudinal and lateral velocity of wheels in wheel or vehicle coordinates\\
		$\psi$, $\dot{\psi}$, $\ddot{\psi}$		&	Yaw angle, rate and acceleration in vehicle's \cg\\
		$l_{\mathrm{f}}$, $l_{\mathrm{r}}$		&	Distance between \cg and front or rear axle\\
		$s_{\mathrm{f}}$, $s_{\mathrm{r}}$		&	Track width front and rear\\
		$\delta_{ij}$, $\dot{\delta}_{ij}$		&	Steering angles and rates of wheels\\
		$\lambda_{ij}$, $\dot{\lambda}_{ij}$	&	Longitudinal slips and rates of wheels\\
		$\alpha_{ij}$, $\dot{\alpha}_{ij}$		&	Slip angles and rates of wheels\\
		$\beta$									&	Side slip angle\\
		$\omega_{ij}$							&	Rotational speed of wheels\\
		$r_{ij}$								&	Effective radius of wheels\\
		$J_{\mathrm{W},ij}$						&	Moment of inertia of wheels\\
		$s$, $\dot s$							&	Traveled distance along path and derivative\\
		$d$, $\dot d$							&	Lateral deviation from path and derivative\\
		\bottomrule
	\end{tabular}
\end{table}

\subsection{Prediction Model Requirements}
\label{subsec:mpcscheme}
Tracking a trajectory in terms of a temporal sequence of reference poses requires taking several aspects into account which determine the prediction model.
First of all, different trajectory implementations are possible. 
For this paper, we assume a trajectory given in Cartesian coordinates plus heading.  
From this, a Frenet description of the trajectory is derived, describing the trajectory in distance $s$ along the path, the lateral displacement $d$, and the yaw angle $\psi$, c.\,f.~\cite{werling_2010}.
The longitudinal and lateral vehicle motion -- as needed for tracking a trajectory -- are part of the prediction model in terms of the variables longitudinal speed $v^{\mathrm{V}}_x$, side slip angle~$\beta$,	 and yaw rate~$\dot{\psi}$. 

The second aspect to be considered is the stability of the vehicle motion. 
In conventional, front axle steered vehicles, the side slip angle serves as measure of stability. 
Vehicle stability control algorithms aim at keeping the side slip angle in defined boundaries. 
In contrast, all-wheel steering enables high side slip angles while the vehicle is stable \cite{russell_2014}. 
Consequently, the wheels' slip angles $\alpha_{ij}$ are considered as measure for vehicle stability and are contained in the prediction model. %

The third aspect to be considered are the actuators' properties and their effects on vehicle dynamics. 
Each actuator is subject to limitations which must be respected in the control law. 
For the steering, these are available steering angles $\delta_{ij}$ and available steering dynamics in terms of steering rates~$\dot{\delta}_{ij}$. 
Brake and drives both effect wheel rotational dynamics.
Thus, these are mutually integrated into the model by the longitudinal slip $\lambda_{ij}$ and its dynamics $\dot{\lambda}_{ij}$.

The last aspect respects the potential of braking with the steering actuators in a configuration comparable to a skier's snowplow brake. 
Instead of braking the wheels, deceleration can be caused through opposing wheel slip angles by turning both wheels of an axle inwards~\cite{reinold_2010,jansen_2010}.
In order to be able to avoid too excessive snowplow braking, the steering angle deviations $\Delta\delta_{\mathrm{f}}$ and $\Delta\delta_{\mathrm{r}}$ between left and right wheel are considered in the prediction model, too.

Summing up these considerations, the used output and control vector for the prediction model result in:
\begin{align*}
	\vc{y} &= \begin{bmatrix}s &d &\psi &v^{\mathrm{V}}_x &\beta &\dot{\psi} &\delta_{ij} &\lambda_{ij} &\alpha_{ij} &\Delta\delta_{i} \end{bmatrix}^T,\\
	\vc{u} &= \begin{bmatrix}\dot{\delta}_{ij} &\dot{\lambda}_{ij} \end{bmatrix}^T.
\end{align*}
Consequently, the resulting prediction model allows to represent the necessary dynamics for trajectory tracking as well as to incorporate of actuator limitations.
The according reference vectors are
\begin{align*}
	\vc{y}_{\mathrm{ref}} &= \begin{bmatrix}s_{\mathrm{ref}} &0 &\psi_{\mathrm{ref}} &v^{\mathrm{V}}_{x,\mathrm{ref}}& 0 &0 &0 &0 &0 &0 \end{bmatrix}^T,\\
	\vc{u}_{\mathrm{ref}} &= \begin{bmatrix}0& 0 \end{bmatrix}^T.
\end{align*}
This reference vector is designed such that the controller tracks the reference distance $s_{\mathrm{\mathrm{ref}}}$, the resulting longitudinal speed $v_{x,\mathrm{\mathrm{ref}}}$, and the reference yaw angle $\psi_{\mathrm{ref}}$. 
All other reference variables are set to zero, targeting a behavior that minimizes actuator utilization as well as lateral vehicle dynamics for passenger comfort.

\subsection{Prediction model}
\label{subsec:vehiclemodel}

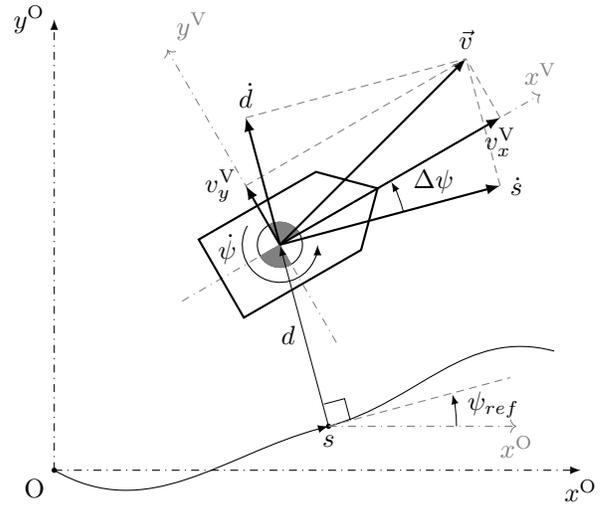
\begin{figure}[]
	\centering
		\begin{tikzpicture}[  
		force/.style={>=latex,thick},
		axis/.style={dash dot,gray},
		speed/.style={thick,black,-latex},
		helpline/.style={densely dashed,gray},
		measurement/.style={>=latex,line width=0.3,shorten >=0.5pt},
		wheel/.style={rectangle,rounded corners=5pt,draw=black,fill=gray!30,minimum height=1cm,minimum width=3cm,thick},
	]
	\def\angledelta{30}
	\def\anglebeta{15}
	\def\vofvehicle{3.5cm}
	\def\arcradius{1.7cm}
	\def\speedwheel{3.5cm}
	\def\anglespeedwheel{17}
	\def\anglepsi{30}
	\def\xshift{3cm}
	\def\yshift{3cm}
	\def\anglepsiref{15}
	\def\frenetd{2.5cm}
	\def\sizerecangle{0.3cm}
	
	\newcommand{\drawcog} {
		\def\radius{0.3cm}
		\def\nsegs{4}
		\def\segangle{360/\nsegs}
		\draw (0,0) circle (\radius);
		\foreach \x in {1,3}
		\filldraw[fill=gray,draw=gray] (0:0mm) -- ({(\x-1)*\segangle}:\radius )
		arc ({(\x-1)*\segangle}:\x*\segangle:\radius ) -- cycle;
		\node[](){};
		{[axis,->]
			\draw (0,-1.5) -- (0,3) node[above right] {$y^{\mathrm{V}}$};
			\draw (-1.5,0) -- (4,0) node[above] {$x^{\mathrm{V}}$};
		}
		\node[](){};
		\begin{scope}[rotate=\anglebeta]
		\draw[speed] (0,0) -- (\vofvehicle,0) node[above] {$\vec{v}$};
		\end{scope}
		\node[](){};
		{[speed]
			\draw (0,0) -- ({\vofvehicle*cos(\anglebeta)},0) node[below]{$v^{\mathrm{V}}_x$};		
			\draw (0,0) -- (0,{\vofvehicle*sin(\anglebeta)}) node[left]{$v^{\mathrm{V}}_y$};
		}
		{[helpline]
			\draw ({\vofvehicle*cos(\anglebeta)},0) -- ({\vofvehicle*cos(\anglebeta)},{\vofvehicle*sin(\anglebeta)});	
			\draw (0,{\vofvehicle*sin(\anglebeta)}) -- ({\vofvehicle*cos(\anglebeta)},{\vofvehicle*sin(\anglebeta)});
		}
		\draw[thick,scale=0.3] (5,0) -- (3,2) -- (-3,2) --(-3,-2)--(3,-2)--(5,0);
		\draw [-latex](120:.5)arc(120:330:.5);
		\node at (155:.7) {$\dot\psi$};	
}

	\node[](){};
	{[axis,->,black]
		\draw[-latex] (0,0) -- (0,6) node[left] {$y^{\mathrm{O}}$};
		\draw[-latex] (0,0) -- (7,0) node[below] {$x^{\mathrm{O}}$};
	}
	\fill (0,0) circle(1pt) node[below left]{$\mathrm{O}$};
	\begin{scope}[yshift=\yshift,xshift=\xshift]
		\coordinate (CoG) at (0,0);
		\begin{scope}[rotate=\anglepsi)]
			\drawcog
		\end{scope}
		\begin{scope}[rotate=\anglepsiref]
			\node[](){};
			{[speed,->]
				\draw (0,0)--({\vofvehicle*cos(\anglebeta+\anglepsiref)},0)[-latex] node[right]{$\dot s$};
				\draw (0,0)--(0,{\vofvehicle*sin(\anglebeta+\anglepsiref)})[-latex]node[above]{$\dot d$};
			}	
			{[helpline]
				\draw ({\vofvehicle*cos(\anglebeta+\anglepsiref)},0) --({\vofvehicle*cos(\anglebeta+\anglepsiref)},{\vofvehicle*sin(\anglebeta+\anglepsiref)});
				\draw (0,{\vofvehicle*sin(\anglebeta+\anglepsiref)}) --({\vofvehicle*cos(\anglebeta+\anglepsiref)},{\vofvehicle*sin(\anglebeta+\anglepsiref)});
			}
			\draw[measurement,->] (0:1*\arcradius)arc(0:\anglepsiref:1*\arcradius);
			\node at (0.5*\anglepsiref:1*1.3*\arcradius) {$\Delta\psi$};	
		\end{scope}
		\coordinate (P) at ({\anglepsiref-90}:\frenetd);
		\fill (P) circle(1pt);
	\end{scope}
	
	\draw (P) to[out=\anglepsiref,in=165] ++(3,1); 
	\draw[<-,>=latex] (P) node[below]{$s$} to[out=180+\anglepsiref,in=-30] (0,0); 
	\draw[helpline] (P) -- ++(\anglepsiref:2.5cm);
	\draw[axis,->](P) -- ++ (2.5cm,0) node[below]{$x^\mathrm{O}$};
	\node[](){};
	{[measurement]
		\draw[->](P)--(CoG) node[left] at ($(P)!0.5!(CoG)$){$d$};
		\begin{scope}[shift={(P)}]]
			\draw[->] (0:1*\arcradius)arc(0:\anglepsiref:1*\arcradius);
			\node at (0.5*\anglepsiref:1.3*\arcradius) {$\psi_{ref}$};
			\draw[rotate=\anglepsiref] (\sizerecangle,0)--(\sizerecangle,\sizerecangle)--(0,\sizerecangle);
		\end{scope}
	}
\end{tikzpicture}
	\caption{Trajectory description in a Frenet representation}
	\label{fig:frenet}
\end{figure}

The vehicle motion along the path $s$ can be described according to \figureref{fig:frenet} as 
\begin{align*}
	\dot{s}&= v^{\mathrm{V}}_x\cos(\psi-\psi_{\mathrm{ref}})-v^{\mathrm{V}}_y\sin(\psi-\psi_{\mathrm{ref}}),\\
	\dot{d}&= v^{\mathrm{V}}_x\sin(\psi-\psi_{\mathrm{ref}})+v^{\mathrm{V}}_y\cos(\psi-\psi_{\mathrm{ref}}).
\end{align*}
The dynamics of lateral and longitudinal tracking relates to the vehicles' actual lateral and longitudinal speeds as well as on the actual yaw angle deviation. 
The dynamics of the yaw angle in terms of the yaw rate $\dot{\psi}$ directly correspond to the vehicles yaw dynamics which are elaborated in the following. 

\begin{figure*}
	\centering
	\def\wheel{\mathrm{W}}
\def\vehicle{\mathrm{V}}
\def\frontright{\mathrm{fr}}
\def\frontleft{\mathrm{fl}}
\def\rearright{\mathrm{rr}}
\def\rearleft{\mathrm{rl}}
\def\rear{\mathrm{r}}
\def\front{\mathrm{f}}

\tikzstyle{force}			 = [>=latex,thick,draw=latForceColor,fill=latForceColor]
\tikzstyle{axis}			 = [dash dot,gray,>=latex]
\tikzstyle{speed}		 = [black,>=latex]
\tikzstyle{helpline}		 = [densely dashed,gray]
\tikzstyle{measurement} = [>=latex,line width=0.3,shorten >=0.5pt]
\tikzstyle{wheel}		 = [rectangle,rounded corners=5pt,draw=black,fill=gray!30,minimum height=1cm,minimum width=3cm,thick]

	\def\angledelta{30}
	\def\anglebeta{15}
	\def\vofvehicle{2.8cm}
	\def\arcradius{1.8cm}
	\def\forcewheel{3.5cm}
	\def\angleforcewheel{45}
	\def\angledelta{30}
	\def\anglebeta{17}
	\def\vofvehicle{3.5cm}
	\def\arcradius{1.7cm}
	\def\speedwheel{3.5cm}
	\def\anglespeedwheel{17}
	\newcommand{\drawwheelForces}[1]{
		\def\wheelnumber{#1}
		\begin{scope}[rotate=\angledelta]
			\node[wheel,transform shape] (drop) {};
			{[axis,->]
				\draw (0,-1) -- (0,2) node[left] {$y^{\wheel}_{#1}$};
				\draw (-2,0) -- (4,0) node[above] {$x^{\wheel}_{#1}$};
			}
			{[force,->]
				\draw (0,0) -- ({\forcewheel*cos(\angleforcewheel-\angledelta)},0) node at ++(-75:13pt) {$F^\wheel_{x,#1}$};		
				\draw (0,0) -- (0,{\forcewheel*sin(\angleforcewheel-\angledelta)}) node[left] {$F^\wheel _{y,#1}$};	
			}
			{[helpline]
				\draw ({\forcewheel*cos(\angleforcewheel-\angledelta)},0) -- ({\forcewheel*cos(\angleforcewheel-\angledelta)},{\forcewheel*sin(\angleforcewheel-\angledelta)});
				\draw (0,{\forcewheel*sin(\angleforcewheel-\angledelta)}) -- ({\forcewheel*cos(\angleforcewheel-\angledelta)},{\forcewheel*sin(\angleforcewheel-\angledelta)});
			}
		\end{scope}
		\node[](){};
		{[axis,->]
			\draw (0,-1) -- (0,3) node[right] {$y^{\vehicle}_{#1}$};
			\draw (-2,0) -- (3.2,0) node[above] {$x^{\vehicle}_{#1}$};
		}
		{[force,->]
			\draw (0,0) -- ({\forcewheel*cos(\angleforcewheel)},0) node[below] {$F^\vehicle_{x,#1}$};		
			\draw (0,0) -- (0,{\forcewheel*sin(\angleforcewheel)}) node[left] {$F^\vehicle_{y,#1}$};	
			\draw (0,0) -- ({\forcewheel*cos(\angleforcewheel)},{\forcewheel*sin(\angleforcewheel)}) node[above] {$F_{#1}$};	
		}
		{[helpline]
			\draw ({\forcewheel*cos(\angleforcewheel)},0) -- ({\forcewheel*cos(\angleforcewheel)},{\forcewheel*sin(\angleforcewheel)});
			\draw (0,{\forcewheel*sin(\angleforcewheel)}) -- ({\forcewheel*cos(\angleforcewheel)},{\forcewheel*sin(\angleforcewheel)});
		}
		\draw[measurement,->] (0:\arcradius)arc(0:\angledelta:\arcradius);
		\node at (0.5*\angledelta:1.2*\arcradius) {$\delta_{#1}$};
	}
	\newcommand{\drawwheelVelocities}[1]{
			\def\wheelnumber{#1}
			\begin{scope}[rotate=\angledelta]
				\node[wheel,transform shape] () {};
				{[axis,->]
					\draw (0,0) -- (0,2) node[left] {$y^{\wheel}_{#1}$};
					\draw (-2,0) -- (4,0) node[above] {$x^{\wheel}_{#1}$};
				}
				{[speed,->]
					\draw (0,0) -- ({\speedwheel*cos(\anglespeedwheel-\angledelta)},0) node at ++(135:18pt) {$v^{\wheel}_{x,#1}$};		
					\draw (0,0) -- (0,{\speedwheel*sin(\anglespeedwheel-\angledelta)}) node at ++(-75:10pt) {$v^{\wheel}_{y,#1}$};	
				}
				{[helpline]
					\draw ({\speedwheel*cos(\anglespeedwheel-\angledelta)},0) -- ({\speedwheel*cos(\anglespeedwheel-\angledelta)},{\speedwheel*sin(\anglespeedwheel-\angledelta)});
					\draw (0,{\speedwheel*sin(\anglespeedwheel-\angledelta)}) -- ({\speedwheel*cos(\anglespeedwheel-\angledelta)},{\speedwheel*sin(\anglespeedwheel-\angledelta)});
				}
			\end{scope}
			\node[](){};
			{[axis,->]
				\draw (0,-1) -- (0,2.5) node[right] {$y^{\vehicle}_{#1}$};
				\draw (-2,0) -- (4,0) node[above left] {$x^{\vehicle}_{#1}$};
			}
			{[speed,->]
				\draw (0,0) -- ({\speedwheel*cos(\anglespeedwheel)},0) node[below] {$v^{\vehicle}_{x,#1}$};		
				\draw (0,0) -- (0,{\speedwheel*sin(\anglespeedwheel)}) node[above right] {$v^{\vehicle}_{y,#1}$};	
				\draw (0,0) -- ({\speedwheel*cos(\anglespeedwheel)},{\speedwheel*sin(\anglespeedwheel)}) node[right] {$\vec{v}_{#1}$};	
			}
			{[helpline]
				\draw ({\speedwheel*cos(\anglespeedwheel)},0) -- ({\speedwheel*cos(\anglespeedwheel)},{\speedwheel*sin(\anglespeedwheel)});
				\draw (0,{\speedwheel*sin(\anglespeedwheel)}) -- ({\speedwheel*cos(\anglespeedwheel)},{\speedwheel*sin(\anglespeedwheel)});
			}
			\draw[measurement,->] (\angledelta:\arcradius)arc(\angledelta:\anglespeedwheel:\arcradius);
			\node at ({0.5*(\angledelta-\anglespeedwheel)+\anglespeedwheel-1}:1.25*\arcradius) {$\alpha_{#1}$};
		}
	\newcommand{\drawcog} {
		\def\radius{0.3cm}
		\def\nsegs{4}
		\def\segangle{360/\nsegs}
		\draw (0,0) circle (\radius);
		\foreach \x in {1,3}
		\filldraw[fill=gray,draw=gray] (0:0mm) -- ({(\x-1)*\segangle}:\radius )
		arc ({(\x-1)*\segangle}:\x*\segangle:\radius ) -- cycle;
		\node at (45:7mm){$m$, $J_z$};
		{[axis,->]
			\draw (0,-1.5) -- (0,3) node[above] {$y^{\vehicle}$};
			\draw (-1.5,0) -- (4,0) node[right] {$x^{\vehicle}$};
		}
		\node[](){};
		{[force,->]
			\draw (0,0) -- ({\forcewheel*cos(\angleforcewheel)},0) node[below left] {$F^{\vehicle}_{x}$, $\dot v^{\vehicle}_x$};		
			\draw (0,0) -- (0,{\forcewheel*sin(\angleforcewheel)}) node[left]  {$F^{\vehicle}_{y}$, $\dot v^{\vehicle}_y$};
			\draw (115:1)arc(115:330:1);
		}
		\begin{scope}[rotate=\anglebeta]
			\draw[speed,->] (0,0) -- (\vofvehicle,0) node[above] {$\vec{v}$};
		\end{scope}
		\node[](){};
		{[speed,->]
			\draw (0,0) -- ({\vofvehicle*cos(\anglebeta)},0) node[below]{$v^{\vehicle}_x$};		
			\draw (0,0) -- (0,{\vofvehicle*sin(\anglebeta)})  node[left]{$v^{\vehicle}_y$};
			\draw (150:1)arc(150:330:1);
			\node at (315:1.5) {$M_z$, $\dot\psi$, $\ddot\psi$};
		}
		{[helpline]
			\draw ({\vofvehicle*cos(\anglebeta)},0) -- ({\vofvehicle*cos(\anglebeta)},{\vofvehicle*sin(\anglebeta)});	
			\draw (0,{\vofvehicle*sin(\anglebeta)}) -- ({\vofvehicle*cos(\anglebeta)},{\vofvehicle*sin(\anglebeta)});
		}	
		\draw[measurement,->] (0:1.5*\arcradius)arc(0:\anglebeta:1.5*\arcradius);
		\node at (0.5*\anglebeta:1.5*1.1*\arcradius) {$\beta$};			
	}
\begin{tikzpicture}
	
	\drawcog
	\begin{scope}[shift={(5cm,3cm)}]
	\drawwheelForces{\frontleft}
	\end{scope}
	\begin{scope}[shift={(5cm,-3cm)}]
	\drawwheelVelocities{\frontright}
	\end{scope}
	\begin{scope}[shift={(-5cm,3cm)}]
	\drawwheelForces{\rearleft}
	\end{scope}
	\begin{scope}[shift={(-5cm,-3cm)}]
	\drawwheelVelocities{\rearright}
	\end{scope}

	\node[](){};
	{[axis]
		\draw (8.3,3) -- ++ (0.6,0) coordinate(vl) -- ++ (0.2,0);
		\draw (8.3,-3) -- ++ (0.6,0) coordinate(vr) -- ++ (0.2,0);
		\draw (-7,3) coordinate(hl) -- ++ (-0.2,0);
		\draw (-7,-3) coordinate(hr) -- ++ (-0.2,0);
		\draw (0,-1.5) -- (0,-4.4)  coordinate(mittel) -- ++(0,-0.2);
		\draw (-5,-4) -- ++ (0,-0.4)  coordinate(hinter) -- ++(0,-0.2);
		\draw (5,-4) -- ++ (0,-0.4)  coordinate(vorder) -- ++(0,-0.2);
	}
	{[measurement,latex-latex]
		\draw (vl) -- (vr) node[right] at ($(vl)!0.5!(vr)$) {$s_\front$};
		\draw (hl) -- (hr) node[right] at ($(hl)!0.5!(hr)$) {$s_\rear$};
		\draw (hinter) -- (mittel) node[above] at ($(hinter)!0.5!(mittel)$) {$l_\rear$};
		\draw (mittel) -- (vorder) node[above] at ($(mittel)!0.5!(vorder)$) {$l_\front$};
	}

\end{tikzpicture}
	\caption{Double track model}
	\vspace{-2em}
	\label{fig:doubletrack}
\end{figure*}
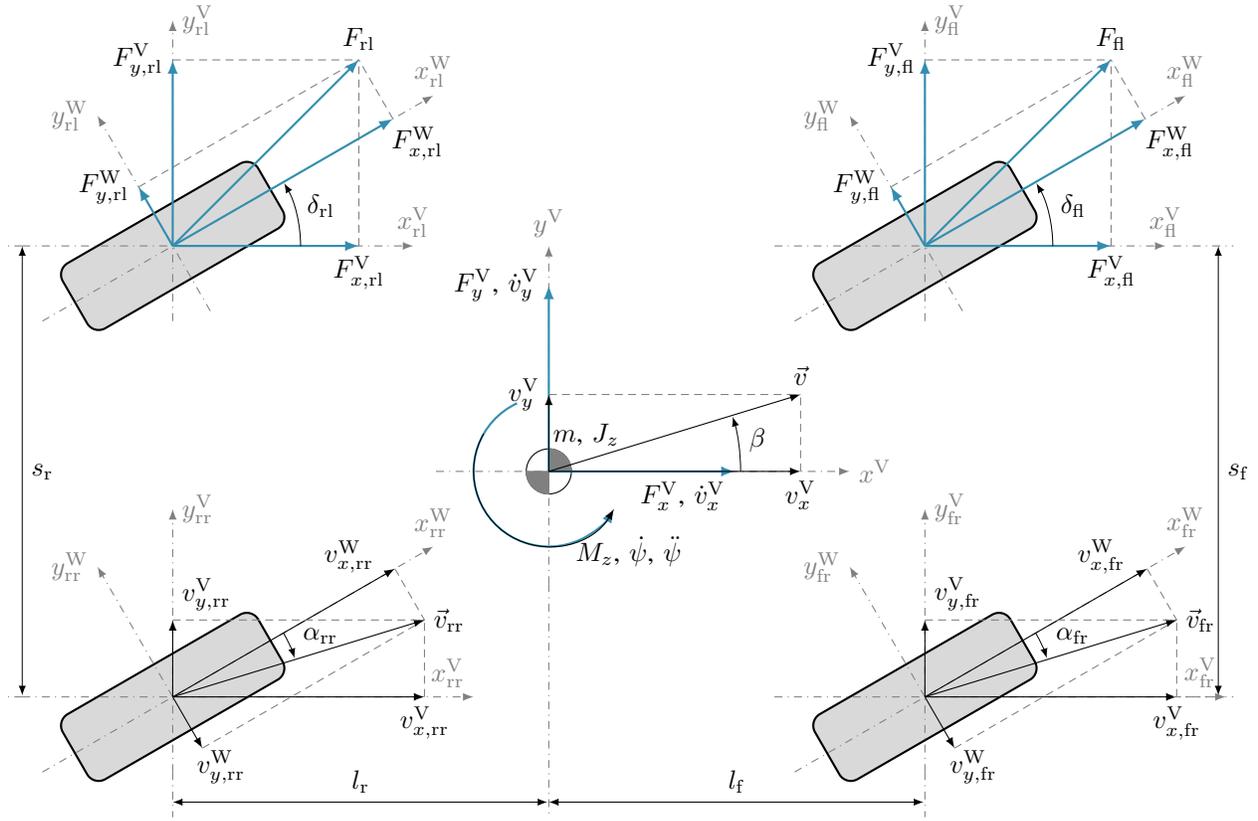

In order to model the influence of independently actuated wheels, a non-linear double track model based on the work of \citet{orend_2006} and \citeauthor{falcone_2007} \cite{falcone_2007,falcone_2007c} is employed together with non-linear tires which combine longitudinal and lateral slip.
Figure \ref{fig:doubletrack} depicts dynamic and kinematic relations in the vehicle model.  
Roll and pitch movements are neglected in the prediction model.

Considering the force and moment equilibria  
with $p_{ij}\in\{\frac{s_{\mathrm{f}}}{2},-\frac{s_{\mathrm{f}}}{2},\frac{s_{\mathrm{r}}}{2},-\frac{s_{\mathrm{r}}}{2}\}$ 
and  $q_{ij}\in\{l_{\mathrm{f}},l_{\mathrm{f}},-l_{\mathrm{r}},-l_{\mathrm{r}}\}$ yields:

\newlength{\whitespace}
\settowidth{\whitespace}{$\:-\:$}
\begin{align*}
\label{eq:equilibria}
\begin{split}
\dot{v}^{\mathrm{V}}_x  = &\hspace{\whitespace}v^{\mathrm{V}}_y\dot{\psi}+\frac{1}{m}\sum_{ij} F^{\mathrm{V}}_{x,ij} \\
= &\hspace{\whitespace}v^{\mathrm{V}}_y\dot{\psi}+\frac{1}{m}\sum_{ij} \left(\cos\delta_{ij} F^{\mathrm{W}}_{x,ij}-\sin\delta_{ij} F^{\mathrm{W}}_{y,ij}\right)\text{,}	
\end{split}
\\
\begin{split}
\dot{v}^{\mathrm{V}}_y  = &-v^{\mathrm{V}}_x\dot{\psi}+\frac{1}{m}\sum_{ij} F^{\mathrm{V}}_{y,ij} \\
= &-v^{\mathrm{V}}_x\dot{\psi}+\frac{1}{m}\sum_{ij} \left(\sin\delta_{ij} F^{\mathrm{W}}_{x,ij}+\cos\delta_{ij} F^{\mathrm{W}}_{y,ij}\right)\text{, }
\end{split}
\\
\begin{split}
\ddot{\psi} = &\hspace{\whitespace}\frac{1}{J_Z}\cdot
\sum_{ij}\left[ \left(p_{ij}\cos\delta_{ij}+q_{ij}\sin\delta_{ij}\right)F^{\mathrm{W}}_{x,ij}\right.\\
&\qquad\qquad\qquad\left.+\left(q_{ij}\cos\delta_{ij}-p_{ij}\sin\delta_{ij}\right)F^{\mathrm{W}}_{y,ij}\right]\text{.}
%
%
%
%
\end{split}
\end{align*}
Consequently, the vehicle motion dynamics depend on the actual vehicle motion and on the forces acting on the wheels. 
Transforming the vehicle's velocity to velocities at the wheel centers yields
\begin{align*}
	v^{\mathrm{V}}_{x,ij}	=	v^{\mathrm{V}}_{x}-p_{ij}\dot{\psi}\text{,\qquad}	
	v^{\mathrm{V}}_{y,ij}	=	v^{\mathrm{V}}_{y}+q_{ij}\dot{\psi}
\end{align*}
and further to wheel coordinates
\begin{align*}
	v^{\mathrm{W}}_{x,ij}	&=	v^{\mathrm{V}}_{x,ij}\cos\delta_{ij}+v^{\mathrm{V}}_{y,ij}\sin\delta_{ij} \text{,}\\	
	v^{\mathrm{W}}_{y,ij}	&=	v^{\mathrm{V}}_{y,ij}\cos\delta_{ij}-v^{\mathrm{V}}_{x,ij}\sin\delta_{ij}.
\end{align*}

Actuator degradations can lead to states in which physical tire limits are completely exploited, 
for instance steering degradations at high lateral accelerations, locking wheels due to brake degradations or undesired anti-lock braking interventions. 
Consequently, modeling the wheel forces $F^{\mathrm{V}}_{x,ij}$ and $F^{\mathrm{V}}_{y,ij}$ requires the representation of the tire's non-linearities as well as the interaction of lateral and longitudinal tire dynamics. 
For this, the Pacejka Magic Formula tire model is employed \cite[pp. 176]{pacejka_2012a} which calculates the longitudinal and lateral tire forces $F^{\mathrm{W}}_{x,ij}$ and $F^{\mathrm{W}}_{y,ij}$ 
as a function of the longitudinal slip $\lambda_{ij}$ and the lateral slip $\alpha_{ij}$. 
For these, following definitions are used:
\begin{align*}
	\lambda_{ij}= 
	\frac{r_{ij}\omega_{ij}-v^{\mathrm{W}}_{x,ij}}{ {\mathrm{max}} (|v^{\mathrm{W}}_{x,ij}|,|r_{ij}\omega_{ij}|)}
	\text{, }
	\alpha_{ij} = \arctan\frac{v^{\mathrm{W}}_{y,ij}}{|v^{\mathrm{W}}_{x,ij}|}\text{.}
\end{align*}

The longitudinal slip $\lambda_{ij}$ is negative for deceleration and positive for acceleration. 
Through utilization of the absolute values in the denominator, the definition is applicable for driving forward as well as reversing. 

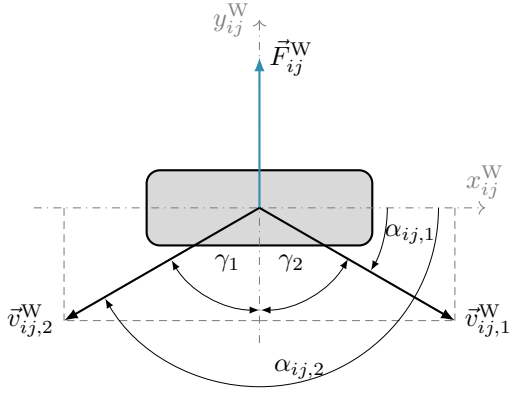
\begin{figure}
	\centering
	\begin{tikzpicture}[  
	force/.style={>=latex,thick,draw=latForceColor,fill=latForceColor},
	axis/.style={dash dot,gray},
	speed/.style={>=latex,thick,black,->},
	helpline/.style={densely dashed,gray},
	measurement/.style={>=latex,line width=0.3,shorten >=0.5pt},
	wheel/.style={rectangle,rounded corners=5pt,draw=black,fill=gray!30,minimum height=1cm,minimum width=3cm,thick},
	]
		
	\def\angledelta{0}
	\def\anglebeta{17}
	\def\vofvehicle{3.5cm}
	\def\arcradius{1.7cm}
	\def\speedwheel{3cm}
	\def\anglespeedwheel{-30}
	\def\forcescale{*0.75}
	\def\fxi{0}
	\def\fyi{4.3073\forcescale}
	\def\fxii{-2.312\forcescale}
	\def\fyii{3.1657\forcescale}
	\def\fxiii{-3.1643\forcescale}
	\def\fyiii{2.0522\forcescale}
	\def\fxiv{-3.268\forcescale}
	\def\fyiv{0.464\forcescale}
	\newcommand{\drawwheel}{
		\begin{scope}[rotate=\angledelta]
			\node[wheel,transform shape] () {};
			{[axis,->]
				\draw (0,-1.8) -- (0,2.5) node[left] {$y^{\mathrm{W}}_{ij}$};
				\draw (-3,0) -- (3,0) node[above] {$x^{\mathrm{W}}_{ij}$};
			}
		
			{[force,->]
			\draw (0,0) -- (0,2) node[right,align=left] {$\vec{F}^{\mathrm{W}}_{ij}$};	
			}
		\end{scope}
		\node[](){};
		{[speed,->]
			\draw (0,0) -- ({\speedwheel*cos(\anglespeedwheel)},{\speedwheel*sin(\anglespeedwheel)}) node[right] {$\vec{v}^{\mathrm{W}}_{ij,1}$};	
		}
		{[helpline]
			\draw ({\speedwheel*cos(\anglespeedwheel)},0) -- ({\speedwheel*cos(\anglespeedwheel)},{\speedwheel*sin(\anglespeedwheel)});
			\draw (0,{\speedwheel*sin(\anglespeedwheel)}) -- ({\speedwheel*cos(\anglespeedwheel)},{\speedwheel*sin(\anglespeedwheel)});
		}
		{[speed,->]
		\draw (0,0) -- (-{\speedwheel*cos(\anglespeedwheel)},{\speedwheel*sin(\anglespeedwheel)}) node[left] {$\vec{v}^{\mathrm{W}}_{ij,2}$};	
		}
	{[helpline]
		\draw (-{\speedwheel*cos(\anglespeedwheel)},0) -- (-{\speedwheel*cos(\anglespeedwheel)},{\speedwheel*sin(\anglespeedwheel)});
		\draw (0,{\speedwheel*sin(\anglespeedwheel)}) -- (-{\speedwheel*cos(\anglespeedwheel)},{\speedwheel*sin(\anglespeedwheel)});
	}
		{[measurement,->] \draw(\angledelta:\arcradius)arc(\angledelta:\anglespeedwheel:\arcradius);
		\node at ({0.7*(\angledelta-\anglespeedwheel)+\anglespeedwheel-1}:1.2*\arcradius) {$\alpha_{ij,1}$};
		\draw(\angledelta:1.4*\arcradius)arc(\angledelta:{-180-\anglespeedwheel}:1.4*\arcradius);
		\node at ({0.5*(\angledelta-(-180-\anglespeedwheel))+(-180-\anglespeedwheel)-1}:1.3*\arcradius) {$\alpha_{ij,2}$};
		
		\draw[latex-latex](\anglespeedwheel:0.8*\arcradius)arc(\anglespeedwheel:-90:0.8*\arcradius);
		\node at ({(\anglespeedwheel)+0.5(-90-\anglespeedwheel)}:0.5*\arcradius) {$\gamma_2$};
		\draw[latex-latex](-90:0.8*\arcradius)arc(-90:-180-\anglespeedwheel:0.8*\arcradius);
		\node at({-90+0.5*(-90-\anglespeedwheel)}:0.5*\arcradius) {$\gamma_1$};
		}	
}
\drawwheel
\end{tikzpicture}
	\caption{Two wheel velocity vectors $\vec{v}^W_{ij,1}$, $\vec{v}^W_{ij,2}$ of same speed and the same resulting lateral force $\vec{F}^W_{ij}$ ($\lambda_{ij}=0$, $\gamma_1=\gamma_2$)}
	\label{fig:alphafres}
	\vspace{-1.5em}
\end{figure}

The same applies to the utilization of the absolute value in the denominator within the definition of the lateral slip~$\alpha_{ij}$. 
Two velocities of the same speed which are symmetric to the wheels lateral axis yield the same resulting lateral force as illustrated in \figureref{fig:alphafres}. 

In order to reach continuously differentiable functions for $\lambda_{ij}$ and $\alpha_{ij}$, the absolute value function is approximated by
\begin{align*}
	|x| = x \text{sgn}(x)\approx x\frac{2}{\pi}\arctan(mx)+n
\end{align*}
where $m$ denotes a parameter to determine the slope of the $\arctan$ around $0$ and $n$ a small offset used to avoid the singularity at $0$. 
We selected $m=5$ and $n=0.1273$. 
The resulting error of this approximation equals $n$ at $0$ and converges to values at the order of $10^{-5}$ for values beyond~$0$.

Finally, as $\beta$ is allowed to reach large values in the context of a front and rear steered vehicle, it is defined as
\begin{align*}
	\beta &= \arctan\frac{v^{\mathrm{V}}_y}{v^{\mathrm{V}}_x}\text{.}
\end{align*}

\subsection{Degradation mapping}
\label{subsec:failuremapping}

For fault-tolerant trajectory tracking, the control scheme must attune to the specific degradation.  
Thus, the considered degradation types and the according mapping in the control scheme are outlined in this subsection. 

Different kinds of functional degradation can occur on actuator level \cite{stolte_2016a}. Degraded brakes or drives can yield
\begin{enumerate}[label={D.\arabic*},leftmargin=3em]
	\item Constant torque at wheel: $M_{\mathrm{W}}=\mathrm{const.}\neq0$\label{deg:torqueconst}
	\item Degraded longitudinal slip range, \eg due to faults in the longitudinal slip control: $\lambda\in [\lambda_{\mathrm{min,deg}}, \lambda_{\mathrm{max,deg}}]$,\label{deg:longsliprange}
	\item Reduced longitudinal slip dynamic, \eg caused by reduced maximum torque at a brake or drive actuator: \newline$\dot{\lambda}\in [\dot{\lambda}_{\mathrm{\min,deg}}, \dot{\lambda}_{\mathrm{max,deg}}]$,\label{deg:longslipdyn}
	\item Zero longitudinal slip, \eg brake or drive do not provide torque: $\lambda=0$,\label{deg:lambda=0}
	\item Constant longitudinal slip, \eg anti-lock control is unnecessarily active: $\lambda=\mathrm{const.}\neq0$, especially $\lambda=\lambda_{\mathrm{\mu,max}}$, \label{deg:lambdaconstant}
	\item Locking or spinning wheel, \eg brake locks wheel or drive provides maximum torque: $\lambda=1$, $\lambda=-1$.\label{deg:lambda1}
\end{enumerate}
For steering following degradation types are possible:
\begin{enumerate}[label={D.\arabic*},leftmargin=3em,start = 7]
	\item Reduced steering angle range: \newline$\delta\in [\delta_{\mathrm{min,deg}}, \delta_{\mathrm{max,deg}}]$,\label{deg:steeringangle}
	\item Reduced steering dynamics: $\dot{\delta}\in [\dot{\delta}_{\mathrm{min,deg}}, \dot{\delta}_{\mathrm{max,deg}}]$,\label{deg:steeringdynamic}
	\item Constant steering angle: $\delta=\mathrm{const.}$, especially \newline$\delta=\delta_{\mathrm{min}}$ and $\delta=\delta_{\mathrm{max}}$,\label{deg:deltaconstant}
	\item Free running steering: Steering behavior is purely influenced by the forces acting at the wheel and determined by the steering kinematics and dynamics (damping etc.).\label{deg:freesteering}
\end{enumerate}

For fault-tolerant control, the degradations are mapped to the control algorithm in the reconfiguration block in \figureref{fig:controlstructure}.
Degradations \ref{deg:longsliprange}, \ref{deg:longslipdyn}, \ref{deg:steeringangle}, and \ref{deg:steeringdynamic} are addressed by adapting the constraints of the allowed value ranges for $\lambda_{ij}$, $\dot{\lambda}_{ij}$, $\delta_{ij}$, and $\dot{\delta}_{ij}$ in $\mathbb{U}$ and $\mathbb{Y}$. 

The remaining degradations with respect to the rotational wheel dynamics \ref{deg:torqueconst}, \ref{deg:lambda=0}, and \ref{deg:lambda1} are addressed by three measures. 
The weight $w_{\lambda,{ij}}$ of the affected actuator is set to zero such that it is not considered in the cost function~$J$. 
Moreover, the constraints $\lambda_{ij,\mathrm{max}}$ and $\lambda_{ij,\mathrm{min}}$ 
are voided in order to prevent problems during optimization as physical slip beyond these values can occur due to the degradation.
The according entries in the transition and input matrices $\mx{A}^{\mathrm{lin}}_{0}$ and $\mx{B}^{\mathrm{lin}}_{0}$ are set to zero, too. 
By this means, the effect of the degraded actuator is considered during the prediction only by the disturbance $\vc{r}_0$. 

The same principle applies to \ref{deg:deltaconstant} by zeroing $w_{\delta,{ij}}$, voiding $\alpha_{ij,\mathrm{max}}$ and $\alpha_{ij,\mathrm{min}}$, as well as zeroing the matrix elements relevant for the affected steering. 
Additionally, voiding $\alpha_{ij,\mathrm{max}}$ and $\alpha_{ij,\mathrm{min}}$ on both wheels of the affected axle allows large slip angles which can be expected during a steering degradation. 
Still, $\alpha_{ij}$ is considered in the cost function $J$ such that the effects of the unconstrained slip angle are minimized with respect to the trajectory tracking. 

A special case is degradation \ref{deg:freesteering}: 
Provided a positive mechanical trail together with longitudinal slip $\lambda_{ij}=0$, the wheel's motion around its $z$-axis follows the vehicle's motion. 
This changes when drive or brake apply a torque. 
In this case, the wheel turns inwards or outwards respectively depending on the scrub radius and the torques direction, thereby magnifying the degradation's effect on the lateral vehicle dynamics. 
Consequently, the measures of \ref{deg:deltaconstant} and \ref{deg:lambda=0} are combined in order to reach a free running wheel without any intentionally applied forces.

\section{Evaluation}
\label{sec:evaluation}
For investigating whether functional redundancies are suitable for ensuring safety on trajectory tracking level, the control algorithm used for trajectory tracking must prove to be capable of handling all degradations accepted by the safety concept.
Furthermore, the degradations can occur during driving manifold trajectories which must be derived from the functional range and the operational design domain of the vehicle automation functionality. 

In this section, we apply the developed control system in an exemplary \emph{sine-with-dwell}-like maneuver at a speed of ca. \SI{14}{\meter\per\second} such that the tire forces are temporarily close to saturation for degradation-free operation, \cf Subsection~\ref{subsec:degradationfree}. 
In Subsection~\ref{subsec:drivebrakedegradations}, the control scheme is applied to selected degradations regarding the rotational wheel dynamics caused by brake and drives while Subsection~\ref{subsec:steeringdegradations} considers steering related degradations.

The control scheme is simulatively evaluated. 
A double track model with roll and pitch dynamics serves as the vehicle plant with the model parameters demonstrated in Table~\ref{tab:modelparameters}. 
A degradation detection and isolation time $T_{\mathrm{DDI}}=\SI{.2}{\second}$  respects the duration of the vehicle's self-diagnosis. 
The degradation is triggered at $t=\SI{1}{\second}$ and a constant available maximum friction $\mu_{\mathrm{max}}=\mathrm{const.}$ is assumed. 
The MPC is running at a sampling time $T_S=\SI{50}{\milli\second}$ with a prediction and control horizon $T_P=\SI{1}{\second}$ and $T_C=\SI{250}{\milli\second}$, respectively.

\begin{table}[b]
	\vspace{-1em}
	\caption{Vehicle Model Parameter Values and Parameter Ranges for Degradation-free Operation}
	\label{tab:modelparameters}
	\centering
	\begin{tabular}{l
			S[table-format=4.2]@{\,}
			s[table-unit-alignment = left]}
		\toprule
		Parameter 								& \multicolumn{2}{c}{Value}\\
		\midrule
		$m$										& 2200	&\kilogram	\\
		$J_z$									& 2000	&\kilogram\square\meter	\\
		$l_{\mathrm{f}}$ 						& 1.36	&\meter\\			
		$l_{\mathrm{r}}$						& 1.36	&\meter	\\
		$s_{\mathrm{f}}$						& 1.75	&\meter	\\ 
		$s_{\mathrm{r}}$						& 1.75	&\meter	\\
		$r_{ij}$								& 0.28	&\meter\\
		$J_{\mathrm{W}}$						& 2		&\kilogram\square\meter \\
		$h_{\mathrm{\cg}}$						& .3	&\meter\\
		\bottomrule
	\end{tabular}	
	\begin{tabular}
		{lc}
		\toprule
		Parameter 			& Value Range\\
		\midrule
		$\delta_{ij}$		& [\SI{-30}{\degree}, \SI{30}{\degree}]\\
		$\dot{\delta}_{ij}$	& [\SI{-120}{\degree\per\second}, \SI{120}{\degree\per\second}]\\
		$M_{\mathrm{W},ij}$		& [\SI{-2000}{\newton\meter}, \SI{2000}{\newton\meter}]\\
		$\lambda_{ij}$		& [\SI{-.12}{}, \SI{.12}{}]\\
		$\alpha_{ij}$		& [\SI{-.2}{}, \SI{.2}{}]\\

		\bottomrule
	\end{tabular}	
	\vspace{-1.5em}
\end{table}

\citet{calzolari_2017} introduce metrics in order to compare different trajectory tracking control approaches, namely the maximum $\varepsilon_{\mathrm{\{t,n\},max}}$, average $\varepsilon_{\mathrm{\{t,n\},avg}}$, and final $\varepsilon_{\{t,n\},\mathrm{end}}$ tangential and normal deviation from the reference trajectory, as well as the average tire saturation $\mu_{\mathrm{avg}}$. 
Thereby, the authors only regard vehicles with front steering. 
However, a vehicle featuring front and rear steering basically allows driven trajectories which track the positions accurately while showing a huge yaw angle deviation. 
Thus, we extend the metrics of \citeauthor{calzolari_2017} by the maximum $\varepsilon_{\mathrm{\psi,max}}$, average $\varepsilon_{\mathrm{\psi,avg}}$, and final yaw angle deviation $\varepsilon_{\mathrm{\psi,end}}$:
\begin{align*}
	\varepsilon_{\mathrm{\psi,max}} &= \underset{t\in[0,T]}{\mathrm{max}} |\psi(t)-\psi_{\mathrm{ref}}(t)|\\
	\varepsilon_{\mathrm{\psi,avg}} &= \frac{1}{T}\int_{0}^{T}|\psi(t)-\psi_{\mathrm{ref}}(t)|dt\\
	\varepsilon_{\mathrm{\psi,end}} &= |\psi(T)-\psi_{\mathrm{ref}}(T)|	
\end{align*}

Table \ref{tab:results} summarizes trajectory tracking performance for the different cases outlined in the following. 
Based on a ride on a major inner city road, tangential deviations $\varepsilon_t>\SI{1}{\meter}$, normal deviations $\varepsilon_n>\SI{0.3}{\meter}$, and a yaw angle deviation  $\varepsilon_\psi>\SI{10}{\degree}$ are assumed as not tolerable from a safety perspective. 
Still, these arbitrary definitions of what is safe must be defined for every individual scenario and, hence, must be further investigated.%

\setlength{\tabcolsep}{0.5\tabcolsep}
\newlength{\metricWidth}
\setlength{\metricWidth}{.9cm}
\newlength{\numberWidth}
\setlength{\numberWidth}{.5cm}
\newlength{\metricWidthHalf}
\setlength{\metricWidthHalf}{.5\metricWidth}
\addtolength{\metricWidthHalf}{-\tabcolsep}
\newlength{\descriptionWidth}
\setlength{\descriptionWidth}{\textwidth}
\addtolength{\descriptionWidth}{-\numberWidth}
\addtolength{\descriptionWidth}{-24.0\tabcolsep}
\addtolength{\descriptionWidth}{-10.0\metricWidth}

\sisetup{
	table-format = 2.2,
	round-precision=2,
	round-mode=places,
}

\pgfplotstableread[col sep=semicolon,header=true]{tables/results_selection.data}\data

\newcommand{\postprocess}{postproc cell content/.style={/pgfplots/table/@cell content/.add={\pgfmathparse{int(greater(##1,1.0))}\ifnum\pgfmathresult=1\boldmath\fi}{}}}

\begin{table*}[h]
	\footnotesize
	\centering
	\caption{Selected control deviations for a sine-with-dwell-like trajectory without degradation (1), with parameter variation (2), and degradations (3--11). Highlighted values are exceeding the assumed safety boundaries.}
	\vspace{-1em}
	\label{tab:results}
	\pgfplotstabletypeset[
		outfile={highlight.tex},
		include outfiles=true,
		columns={
			Number,
			Description,
			ETmax,
			ETavg,
			ETend, 
			ENmax, 
			ENavg,
			ENend, 
			PSImax, 
			PSIavg, 
			PSIend, 
			MUavg
		},
		display columns/0/.style=  {column type= {c}, string type, column name ={\#}},
		display columns/1/.style=  {column type= {l}, string type, column name ={Description}},
		display columns/2/.style=  {dec sep align={r}, precision=2, fixed, fixed zerofill, column name = {\parbox{\metricWidth}{\centering$\varepsilon_{\mathrm{t,max}}$ \newline in \si{\meter}}},%
									postproc cell content/.style={/pgfplots/table/@cell content/.add={\pgfmathparse{int(greater(##1,1.0))}\ifnum\pgfmathresult=1\boldmath\fi}{}}},	
		display columns/3/.style=  {dec sep align={c}, precision=2, fixed, fixed zerofill, column name = {\parbox{\metricWidth}{\centering$\varepsilon_{\mathrm{t,avg}}$ \newline in \si{\meter}}},
			postproc cell content/.style={/pgfplots/table/@cell content/.add={\pgfmathparse{int(greater(##1,1.0))}\ifnum\pgfmathresult=1\boldmath\fi}{}}},	 
		display columns/4/.style=  {dec sep align={c}, precision=2, fixed, fixed zerofill, column name = {\parbox{\metricWidth}{\centering$\varepsilon_{\mathrm{t,end}}$ \newline in \si{\meter}}}, postproc cell content/.style={/pgfplots/table/@cell content/.add={\pgfmathparse{int(greater(##1,1.0))}\ifnum\pgfmathresult=1\boldmath\fi}{}}},	
		display columns/5/.style=  {dec sep align={c}, precision=2, fixed, fixed zerofill, column name = {\parbox{\metricWidth}{\centering$\varepsilon_{\mathrm{n,max}}$ \newline in \si{\meter}}}, postproc cell content/.style={/pgfplots/table/@cell content/.add={\pgfmathparse{int(greater(##1,0.3))}\ifnum\pgfmathresult=1\boldmath\fi}{}}},	
		display columns/6/.style=  {dec sep align={c}, precision=2, fixed, fixed zerofill, column name = {\parbox{\metricWidth}{\centering$\varepsilon_{\mathrm{n,avg}}$ \newline in \si{\meter}}}, postproc cell content/.style={/pgfplots/table/@cell content/.add={\pgfmathparse{int(greater(##1,0.3))}\ifnum\pgfmathresult=1\boldmath\fi}{}}},	
		display columns/7/.style=  {dec sep align={c}, precision=2, fixed, fixed zerofill, column name = {\parbox{\metricWidth}{\centering$\varepsilon_{\mathrm{n,end}}$ \newline in \si{\meter}}}, postproc cell content/.style={/pgfplots/table/@cell content/.add={\pgfmathparse{int(greater(##1,0.3))}\ifnum\pgfmathresult=1\boldmath\fi}{}}},	
		display columns/8/.style=  {dec sep align={c}, precision=2, fixed, fixed zerofill, column name = {\parbox{\metricWidth}{\centering$\varepsilon_{\mathrm{\psi,max}}$ \newline in \si{\degree}}}, postproc cell content/.style={/pgfplots/table/@cell content/.add={\pgfmathparse{int(greater(##1,10.0))}\ifnum\pgfmathresult=1\boldmath\fi}{}}},	
		display columns/9/.style=  {dec sep align={c}, precision=2, fixed, fixed zerofill, column name = {\parbox{\metricWidth}{\centering$\varepsilon_{\mathrm{\psi,avg}}$ \newline in \si{\degree}}}, postproc cell content/.style={/pgfplots/table/@cell content/.add={\pgfmathparse{int(greater(##1,10.0))}\ifnum\pgfmathresult=1\boldmath\fi}{}}},	
		display columns/10/.style= {dec sep align={c}, precision=2, fixed, fixed zerofill, column name = {\parbox{\metricWidth}{\centering$\varepsilon_{\mathrm{\psi,end}}$ \newline in \si{\degree}}}, postproc cell content/.style={/pgfplots/table/@cell content/.add={\pgfmathparse{int(greater(##1,10.0))}\ifnum\pgfmathresult=1\boldmath\fi}{}}},	
		display columns/11/.style= {dec sep align={c}, precision=2, fixed, fixed zerofill, column name = {\parbox{\metricWidth}{\centering$\mathrm{\mu_{avg}}$}}},%
		every head row/.style={before row=\toprule,
							   after row=\midrule},
		every last row/.style={after row=\bottomrule},
	]
{\data}
\end{table*}

\subsection{Degradation-free Operation}
\label{subsec:degradationfree}
\begin{figure*}
	\centering
	\input{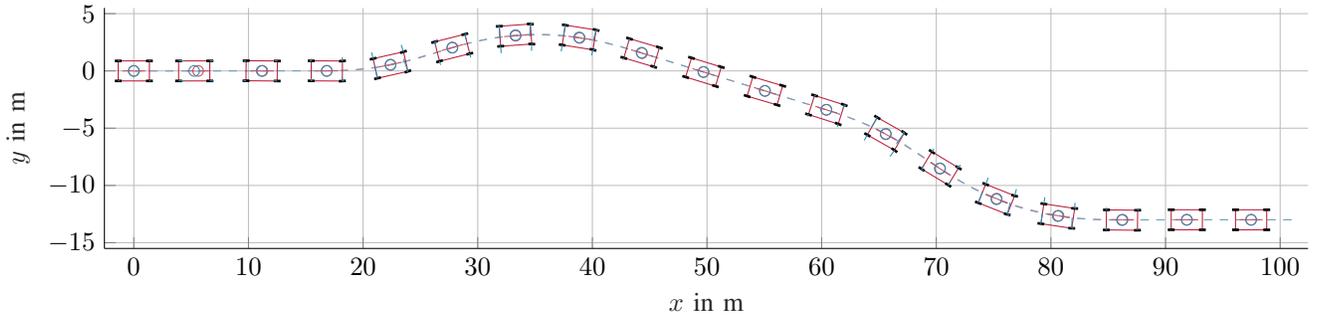}
	\vspace{-1em}
	\caption{Tracked trajectory without degradation, (1) in Table \ref{tab:results}}
	\label{fig:degradationfree}
	\vspace{-1.em}
\end{figure*}

Targeting the general suitability of the proposed controller for trajectory tracking, the maneuver is driven without any degradations (1) as depicted in \figureref{fig:degradationfree}. 
The results indicate a trajectory tracking performance of the same order as other trajectory tracking controllers~\cite{calzolari_2017}, albeit different trajectories are used.
The maximum tangential deviation results from longitudinal speed error at the simulation start. 
The tangential speed is initialized as $v^V_{x,0}=\SI{12}{\meter\per\second}$ while the reference speed is $v^V_{x,\mathrm{ref},0}=\SI{14}{\meter\per\second}$.
Consequently, the vehicle accelerates at the beginning and compensates for the resulting position error.

In order to investigate the robustness of the control approach, in (2) the plant's mass $m$ and yaw inertia $J_z$ are increased by \SI{10}{\percent} each while the \cg of the plant is moved \SI{.2}{\meter} to the rear. 
In this case, no significant changes of the metrics are observable. 

\subsection{Drive and Brake Degradations}
\label{subsec:drivebrakedegradations}
Brake and Drive degradations are considered together since both affect the wheel's rotational dynamics. 
Different types of degradation with varying parameterization of the degradations were investigated. 
Table \ref{tab:results} contains selected results for degradation types \ref{deg:torqueconst}, \ref{deg:lambda=0}, \ref{deg:lambdaconstant}, and \ref{deg:lambda1}.

When applying a constant torque to one wheel \ref{deg:torqueconst} -- such that the resulting longitudinal slip is within the linear tire region -- the effects are compensated by the controller~(3). 
In this case, the controller primarily applies a negative torque of the same order to the other wheel of the same side of the vehicle.
In consequence, also degradation \ref{deg:lambda=0} representing zero longitudinal slip is easily compensated~(4). 

\begin{figure*}
	\centering
	\input{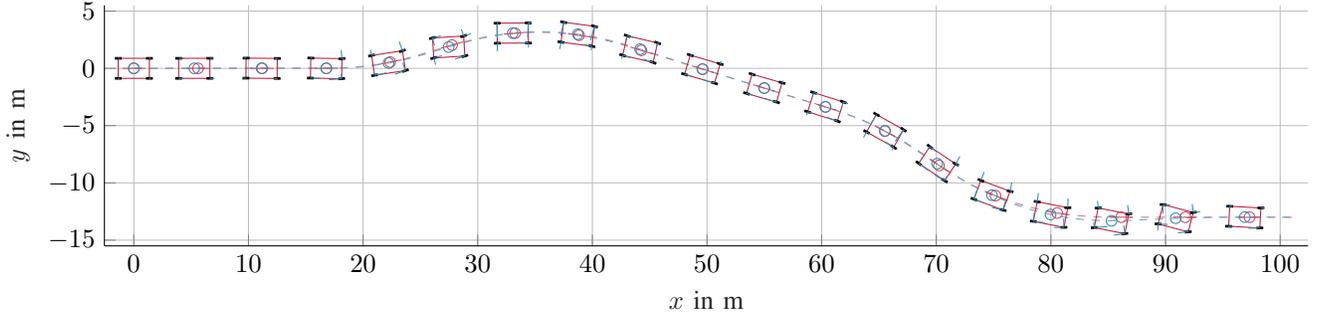}
	\vspace{-1em}
	\caption{Tracked trajectory with undesired front right anti-lock braking $\lambda_{\mathrm{fr}}=\SI{-0.13}{}=\mathrm{const.}$, (5) in Table \ref{tab:results}}
	\label{fig:withdegradation2}
	\vspace{-1em}
\end{figure*}

Degradation types \ref{deg:lambdaconstant} and \ref{deg:lambda1} are more critical, which represent an unnecessarily active anti-lock control (5) or a locking wheel~(6), respectively. 
The results imply that more lateral force potential is left when anti-lock control is active (see \figureref{fig:withdegradation2}) compared to a locking wheel. 
In contrast, the locking wheel yields the most severe position deviation for the trajectory and the used control algorithm. 
Moreover, the yaw angle deviation is comparably high, too. 

\subsection{Steering Degradations}
\label{subsec:steeringdegradations}
With respect to steering degradations, degradation types \ref{deg:steeringangle}, \ref{deg:steeringdynamic}, and \ref{deg:deltaconstant} could be investigated as the controlled vehicle model does not yet feature a comprehensive representation of the steering dynamics including kinematics for consideration of a free running steering (\ref{deg:freesteering}).

Degradation types \ref{deg:steeringangle} (7) and \ref{deg:steeringdynamic} (8) do not show a severe impact on the tracked trajectory. 
The yaw angle tracking performance is slightly, yet not significantly, decreased compared to the degradation-free operation. 
In contrast, degradation type \ref{deg:deltaconstant} has a higher impact on trajectory tracking. 
The tracking performance deteriorates for increasing absolute values of $|\delta_{ij}|\!=\!\mathrm{const.}$  
While the effects manifest in a yaw angle deviation in particular, the effects vary depending on the affected wheel and the turn direction of the steering angle. 
For $\delta_{ij}\!=\!0\!=\!\mathrm{const.}$ (9), the tracking performance goes beyond the tolerated yaw angle deviation when affecting the front right wheel. 
In case of $|\delta_{ij}|\!=\!\SI{5}{\degree}\!=\!\mathrm{const.}$ (10), the degradation's effects on each front wheel are critical. 
The observed most severe yaw angle deviation is encountered at $\delta_{\mathrm{fr}}\!=\!\SI{-30}{\degree}\!=\!\mathrm{const.}$ (11) as illustrated in \figureref{fig:withdegradation}.

\begin{figure*}
	\centering
	\input{figures/withdegradation.tikz}
	\vspace{-1em}
	\caption{Tracked trajectory with degraded front right steering angle $\delta_{\mathrm{fr}}=\SI{-30}{\degree}=\mathrm{const.}$, (11) in Table \ref{tab:results}}
	\label{fig:withdegradation}
	\vspace{-1em}
\end{figure*}

\section{Conclusion and Outlook}
\label{sec:summary}
Through the developed fault-tolerant control approach, we are able to investigate the effects of different actuator degradations on tracking a reference trajectory. 
From a trajectory tracking perspective, the results suggest that functional redundancies can be exploited  if certain degradation types are excluded by means of a safety concept. 
Degradations which lead to an operation of the actuators at their physical limits must be avoided in particular. 
For drives and brakes, a considerable safe state is to apply zero or a small torque as these can be coped with by the control algorithm. 
Simultaneously, these degradations must be indicated to the fault-tolerant control algorithm. 
In contrast, a steering actuator must be designed as fail-operational actuator. 
Even minor degradations such as a small constant steering angle lead to signification deviations from the reference trajectory. 

Still, the results are indications only and require further research before functional redundancies can be exploited. 
On the one hand, only a single trajectory is investigated.
Thus, a set of trajectories must be derived from the automated vehicle's operational ranges and its operational design domain. 
On the other hand, the investigations are made without noise and with limited model uncertainties. 
Moreover, superimposed architectural levels could contribute to balance shortcomings on the trajectory tracking level by concurrent adaptation to the degradation. 
Last but not least, the argumentation is made from a safety engineering's perspective and does not respect other disciplines, \eg availability engineering or ergonomics. 

Our future research can be divided in three threads. 
The first thread consists of improving the simulation capabilities by extending the vehicle model, continuing the presented investigations towards different trajectories, varying degradation detection and isolation times, as well as model and sensor noise. 
The second thread is connecting fault-tolerant trajectory tracking with additional architectural levels. 
This is highly relevant for a statement whether functional redundancies can be used. 
In this regard, combining trajectory tracking with fault-tolerant trajectory generation as proposed by \citet{nolte_2017} is the first step. 
The last thread is implementing the control approach into our experimental vehicle \textsc{Mobile} \cite{bergmiller_2015} in order to do real-world experiments.

\section*{Acknowledgment}
The authors would like to thank Christina Harms for proofreading our work.

\renewcommand*{\bibfont}{\footnotesize} 
\printbibliography 
\end{document}